\newif\iftechrep
\techreptrue

\documentclass[conference]{IEEEtran}
\IEEEoverridecommandlockouts

\usepackage{floatrow}
\usepackage{floatflt}

\usepackage{verbatim} 
\usepackage{amssymb}
\usepackage{lipsum}
\usepackage{multicol}
\usepackage{amsmath}
\usepackage{amsthm}
\usepackage{color}
\usepackage{framed} 
\usepackage{hyperref}
\hypersetup{
    colorlinks = true,
    linkcolor = {red},
    urlcolor = {blue},
    citecolor = {blue},
}
\usepackage{macros}
\usepackage{url}
\usepackage{xspace}
\usepackage{xifthen}
\usepackage{listings}
\usepackage{graphicx}
\usepackage{cite}
\usepackage{paralist}
\usepackage{comment}
\usepackage{parcolumns}
\usepackage{fancyvrb}
\usepackage{nicefrac}
\usepackage[draft]{fixme}
\usepackage[inline]{enumitem}
\setlist{nolistsep}
\newif\ifcode
\codefalse
\codetrue
\ifcode
\usepackage[ruled]{algorithm}
\usepackage{algpseudocode}
\usepackage{ioa_code}

\renewcommand{\floatpagefraction}{1}

\usepackage{tikz}

\usepackage{cleveref}
\crefname{section}{\S}{\S\S}
\newcommand{\RNum}[1]{\uppercase\expandafter{\romannumeral #1\relax}}

\newenvironment{proofsketch}[1][Proof sketch]{\noindent\textbf{#1.} }{\hfill $\Box$\\[2mm]}

\def\P{\ensuremath{\mathcal{P}}}

\def\S{\ensuremath{\mathcal{S}}}

\newcommand{\true}{\mathit{true}}
\newcommand{\false}{\mathit{false}}

\newcommand{\remove}[1]{}

\newcommand{\ignore}[1]{}

\newcommand{\LL}{\ms{LL}}
\newcommand{\name}{\texttt{VBL}\xspace}

\definecolor{dkgreen}{rgb}{0,0.6,0}
\definecolor{gray}{rgb}{0.5,0.5,0.5}
\definecolor{mauve}{rgb}{0.58,0,0.82}

\VerbatimFootnotes

\definecolor{gpcolor}{rgb}{0.6,0.2,0.3}
\newboolean{showcomments}
\setboolean{showcomments}{true}
\ifthenelse{\boolean{showcomments}}
{ \newcommand{\mynote}[3]{
    \fbox{\bfseries\sffamily\scriptsize#1}
    {\small$\blacktriangleright$\textsf{\emph{\color{#3}{#2}}}$\blacktriangleleft$}}
}
{
\newcommand{\mynote}[3]{}}

\newcommand{\myparagraph}[1]{\vspace{2.5mm}\noindent\textbf{{#1}.}}

\remove{
\usepackage[compact]{titlesec}
\titlespacing*{\section}{0pt}{*.40}{*.40}
\titlespacing*{\subsection}{0pt}{*.30}{*.30}
\titlespacing*{\subsubsection}{0pt}{*.40}{*.40}
\titlespacing*{\paragraph}{0pt}{*.40}{*.40}
}

\newtheorem{definition}{Definition}
\newtheorem{theorem}{Theorem}[section]

\begin{document}

\hyphenation{con-cur-ren-cy}
\hyphenation{con-cur-ren-cy-op-ti-mal}

\bibliographystyle{splncs}

\title{Optimal Concurrency for List-Based Sets}

\makeatletter
\newcommand{\linebreakand}{%
  \end{@IEEEauthorhalign}
  \hfill\mbox{}\par
  \mbox{}\hfill\begin{@IEEEauthorhalign}
}
\makeatother

\author{\IEEEauthorblockN{Vitaly Aksenov}
\IEEEauthorblockA{\textit{ITMO University}\\
Saint-Petersburg, Russia}
\and
\IEEEauthorblockN{Vincent Gramoli}
\IEEEauthorblockA{\textit{University of Sydney and EPFL}\\
Lausanne, Switzerland}
\and
\IEEEauthorblockN{Petr Kuznetsov}
\IEEEauthorblockA{\textit{LTCI, T\'el\'ecom Paris}\\
\textit{Institut Polytechnique Paris} \\
Paris, France}
\linebreakand
\IEEEauthorblockN{Srivatsan Ravi}
\IEEEauthorblockA{
\textit{University of Southern California}\\
Los Angeles, USA}
\and
\IEEEauthorblockN{Di Shang}
\IEEEauthorblockA{\textit{University of Sydney}\\
Sydney, Australia}
}

\maketitle
\thispagestyle{plain}
\pagestyle{plain}

\ignore{
\titlerunning{Extracting Maximum Concurrency from List-Based Sets}
\toctitle{Extracting Maximum Concurrency from List-Based Sets}
\author{Vitaly Aksenov\inst{1} \and Vincent Gramoli\inst{2} \and Petr Kuznetsov\inst{3} \and Srivatsan Ravi\inst{4} \and Di Shang\inst{5}}
\authorrunning{V. Aksenov \and V. Gramoli \and P. Kuznetsov \and S. Ravi \and D. Shang}

\institute{
  INRIA Paris / ITMO University
\and 
  University of Sydney
\and 
  LTCI, T\'el\'ecom Paris, Institut Polytechnique Paris
\and
  University of Southern California 
  \and
  IBM
\date{}}
}


\maketitle

\begin{abstract}
Designing an efficient concurrent data structure is an important challenge
that is not easy to meet. 
Intuitively, efficiency of an implementation is defined, in the first place, by its ability to process applied operations \emph{in parallel}, without using unnecessary synchronization. 
As we show in this paper, even for a data structure as simple as a
linked list used to implement the \emph{set} type, the most efficient algorithms known so far are \emph{not concurrency-optimal}: they may reject correct concurrent schedules. 

We propose a new algorithm for the list-based set
based on a \emph{value-aware try-lock} 
that we show to achieve \emph{optimal} concurrency:
it only rejects concurrent schedules that violate correctness of
the implemented set type.
We show empirically that 
reaching optimality does not induce a
significant overhead. In fact, our implementation of the concurrency-optimal
algorithm outperforms both the Lazy Linked List and the Harris-Michael state-of-the-art algorithms.
\end{abstract}
\section{Introduction}\label{sec:intro}
Multicore applications require
highly concurrent data structures.
Yet, the very notion of concurrency is vaguely defined, to say the
least. 
What do we mean by a ``highly  
concurrent'' data structure?  
%
Generally speaking, one could compare the concurrency of algorithms by
running a game where the adversary decides on the schedules of shared memory accesses
from different processes. At the end of the game, the more schedules
the algorithm would accept without hampering high-level correctness,  the more concurrent it would be.
The algorithm that accepts all correct schedules would then be
considered \emph{concurrency-optimal}.

\sloppy{To illustrate the difficulty of optimizing
concurrency,
let us consider one of the most
``concurrency-friendly'' data structures~\cite{Sut08}: the sorted linked list used to 
implement the integer set type.
Since any modification on a linked list affects only a small number
of contiguous list nodes, most of update operations 
on the list 
could, 
in principle, run concurrently without conflicts. 
%
%
%
For example, one of the most efficient concurrent list-based set to
date, the Lazy Linked List~\cite{HHL+05},
achieves high concurrency 
by holding locks on only two consecutive nodes when updating, 
thus accepting modifications of non contiguous nodes 
to be scheduled in any order.
The Lazy Linked List is known to outperform the Java variant~\cite{HS12-book} of the CAS-based
Harris-Michael algorithm~\cite{harris-set,michael-set} 
under low contention because all its traversals, be they for 
read-only look-ups or for locating the nodes to be updated, are \emph{wait-free}, \emph{i.e.}, they ignore
locks and logical deletion marks.
As we show below, the Lazy Linked List implementation is however not
concurrency-optimal, raising two questions: Is there a more concurrent
list-based set algorithm? And if so, does higher concurrency 
induce an overhead that precludes higher performance?
}

The concurrency limitation of the Lazy Linked List is caused by the locking
strategy of its update operations: 
both $\lit{insert}(v)$ and $\lit{remove}(v)$ traverse the structure
until they find a node whose value is larger or equal to $v$, at which
point they acquire locks on two consecutive nodes. Only then is the
existence of the value $v$ checked: 
if $v$ is found (resp. not found), then the insertion (resp., removal)
releases the locks and returns without modifying the structure.
By modifying metadata during lock acquisition without necessarily
modifying the structure itself, the Lazy Linked List over conservatively rejects 
certain correct schedules.
To illustrate that the concurrency limitation of the Lazy Linked List may lead to poor scalability, consider Figure~\ref{fig:lll-contention} that depicts the
performance of a $25$-node Lazy Linked List (red curve) under a workload of $20$\% updates ($\lit{insert}$/$\lit{removals}$) and $80$\% $\lit{contains}$ on a $72$-core machine.
The list is comparatively small, hence all updates (even the failed insertions and
removals) are likely to contend. 
We can see that when we increase the number of threads beyond $40$, the performance drops significantly.
This observation suggests a desirable property 
that concurrent operations should conflict on
metadata only when they conflict on data, for which we need to exploit the semantics of the high-level data type. 
Note that this property refines the original notions of disjoint access
parallelism (DAP)~\cite{tm-book},
trivially ensured by most linked-list implementations simply
because all their operations ``access'' the \textit{head} node and, thus,
are allowed to conflict on the metadata.

\begin{figure}[t]
\hspace{-2em}\includegraphics[scale=0.8]{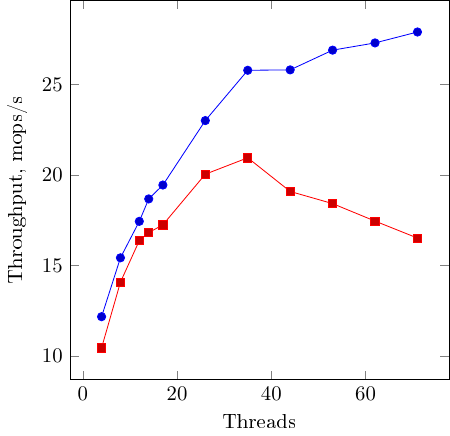}
\caption{\small{The throughput of Lazy Linked List (red square curve) and \name (blue circle curve). We consider the load with only 20\% updates.
  Lazy Linked List behaves worse, as its operations potentially contend on meta-data even when they do not modify the data structure.}
  \label{fig:lll-contention}}
\end{figure}

Our main contribution is the Value-Based List (\texttt{VBL}), the most
concurrent 
(in fact, \emph{optimally} concurrent, as we formally prove)
and probably the most efficient list-based set algorithm to date.
It exploits the logical deletion technique of Harris-Michael that divides the removal of a node into a logical step (marking the node for deletion) and a physical step (unlinking the node from the list),
and the wait-free traversal of the Lazy Linked List.
In addition, our approach relies on a novel \emph{value-aware} synchronization technique:
 first the lock, implemented using \emph{compare-and-swap}, is taken, then the procedure checks whether the \emph{value} in the next node has changed,
 if the validation is successful then the operation continues, otherwise, the operation restarts.
 Compared to the Lazy Linked List, this approach allows for the improvement of performance and even provides scalability in the highly contended cases (Figure~\ref{fig:lll-contention}).
We show that the resulting algorithm rejects a concurrent
schedule only if otherwise the high-level correctness of the
implemented \textsf{set} type (linearizability~\cite{HW90}) is violated. 
Our algorithm is thus
concurrency-optimal: no correct list-based set algorithm can
accept more schedules.

The evaluation of \name shows
that achieving optimal concurrency does not necessary result in
a costly overhead.
Extensive experiments on two x86-64 architectures machines, $72$-way Intel machine and $64$-way AMD machine, confirmed
that \name outperforms the state-of-the-art algorithms~\cite{HHL+05,HS12-book}.
In particular, \name outperforms 
the Lazy Linked List performance 
by $1.6\times$ for $72$ threads on the $20$\%-update workload of Figure~\ref{fig:lll-contention}, which can be explained by the fact that our algorithm validates list data \emph{before} locking, and not after. 
In addition,
as our algorithm differs from Harris-Michael by avoiding metadata accesses during traversals,
it outperforms it by up to $1.6\times$ on read-only workloads.

We report the performance of the Java variant of Harris-Michael list-based set with wait-free $\lit{contains}$
as presented in Shavit and Herlihy's book~\cite{HS12-book} 
and the Java optimised implementation with RTTI~\cite{HHL+05}, 
and, in the technical report~\cite{AGK20},on the performance of our own C++ translations of the Lazy algorithm (without memory management). 

\myparagraph{Roadmap}
The rest of this paper is structured as follows. 
We present our methodology on modelling concurrency and prove the suboptimal concurrency of the Lazy and Harris-Michael linked lists in Section~\ref{sec:concur}. In Section~\ref{sec:algolist}, we present our \name list implementation. 
Section~\ref{sec:eval} presents the methodology for performance evaluation of concurrent list implementations and Section~\ref{sec:conclusion} presents a discussion of concurrency w.r.t list-based sets. 
\iftechrep
 The optional appendix contains the full proofs of linearizability and deadlock-freedom. 
\else 
 The full proofs of linearizability and deadlock-freedom are deferred to the technical report~\cite{AGK20}.
\fi 
Synchrobench benchmark suite~\cite{Gra15} contains the code for all the lists considered in this paper.

%
%
\section{Concurrency analysis of list-based sets}\label{sec:concur}
%
%
\subsection{Preliminaries}
We consider a standard asynchronous shared-memory system, in which $n>1$
processes (or \emph{threads} of computation) $p_1,\ldots , p_n$ communicate by applying operations on
shared \emph{objects}. 

\subsection{Sequential list-based set}

An \emph{abstract data type} $\tau$ is a tuple
$(\Phi,\Gamma, Q, q_0, \delta)$ where
$\Phi$ is a set of operations,
$\Gamma$ is a set of responses, $Q$ is a set of states, $q_0\in Q$ is an
initial state and 
$\delta \subseteq Q\times \Phi \times Q\times \Gamma$ 
is a transition relation that determines, for each state
and each operation, the set of possible
resulting states and produced responses~\cite{AFHHT07}. 
Here, $(q,\pi,q',r) \in \delta$ implies that when
an operation $\pi \in \Phi$ is applied on an object of type $\tau$
in state $q$, the object may move to state $q'$ and return a response $r$.

An object of the \emph{set} type stores a set of integer values,
initially empty, and exports
operations $\lit{insert}(v)$, $\lit{remove}(v)$, $\lit{contains}(v)$ where $v \in \mathbb{Z}$.
The update operations, $\lit{insert}(v)$ and $\lit{remove}(v)$, return
a boolean response, $\true$ if and only if $v$ is absent (for
$\lit{insert}(v)$) or present (for $\lit{remove}(v)$) in the list.  
After $\lit{insert}(v)$ is complete, $v$ is present in the list, and 
after $\lit{remove}(v)$ is complete, $v$ is absent from the list.
The $\lit{contains}(v)$ returns a boolean $\true$ if and
only if $v$ is present in the list.

\begin{algorithm*}[ht!]
\caption{Sequential implementation {\LL} (\textit{sorted linked list}) of \emph{set} type: Shared memory reads and writes are explicitly depicted}
\label{alg:lists}
  \begin{algorithmic}[1]
  	\begin{multicols}{2}
  	{\footnotesize
	\Part{Shared variables}{
		\State $\textbf{class Node}$: $\ms{head}$, $\ms{tail}$
		\State ~~~$\ms{head}.val=-\infty$
		\State ~~~$\ms{tail}.val=+\infty$
		\State ~~~$\ms{head}.next=\ms{tail}$
	}\EndPart
	
	\Statex
	
	\Part{$\lit{insert}(v$)}{
		\State $\ms{prev} \gets \ms{head}$  			                
		\State $\ms{curr} \gets \lit{read}(\ms{prev.next})$ 		
		\While{$(\ms{tval} \gets \lit{read}(\ms{curr.val})) < v $}        
			\State $\ms{prev} \gets \ms{curr}$ 				
			\State $\ms{curr} \gets \lit{read}(\ms{curr.next})$ 	
		\EndWhile
		\If{$\ms{tval} \neq v$}							
			\State $X \gets \lit{new-node}(v,\ms{prev.next})$ 	\label{line:newnode}
                        \State $\lit{write}(\ms{prev.next}, X)$ 	    \label{line:seqinswrite}          
		\EndIf
		\Return $(\ms{tval}\neq v)$ 						
		\EndReturn
   	}\EndPart
	
	\newpage
	
	\Part{$\lit{remove}(v$)}{

		\State $\ms{prev} \gets \ms{head}$  			
		\State $\ms{curr} \gets \lit{read}(\ms{prev.next})$ 		
		\While{$(\ms{tval} \gets \lit{read}(\ms{curr.val})) < v $} 	
					
			\State $\ms{prev} \gets \ms{curr}$ 			
			\State $\ms{curr} \gets \lit{read}(\ms{curr.next})$ 	
		\EndWhile
		\If{$\ms{tval} = v$}							
                        \State $tnext \gets \lit{read}(\ms{curr.next})$  
			\State $\lit{write}(\ms{prev.next}, tnext)$   
		\EndIf
		\Return $(\ms{tval}=v)$ 						
		\EndReturn	
   	}\EndPart
	
	\Statex
	
	\Part{$\lit{contains}(v$)}{
	  \State $\ms{curr} \gets \ms{head}$  			            
		\State $\ms{curr} \gets \lit{read}(\ms{prev.next})$ 		
		\While{$(\ms{tval} \gets \lit{read}(\ms{curr.val})) < v $} 	
			\State $\ms{curr} \gets \lit{read}(\ms{curr.next})$ 	
		\EndWhile
 	   	\Return $(\ms{tval}=v)$						
 	   	\EndReturn
   	 }\EndPart		
	}
	\end{multicols}
  \end{algorithmic}
\end{algorithm*}

The sequential implementation \LL~of the \emph{set} type is
presented in Algorithm~\ref{alg:lists}. The implementation uses a \emph{sorted linked list} data structure 
in which each node (except the \emph{tail}) maintains a \textit{next} field to provide a pointer to the
successor node. Initially, the \emph{next} field of the \emph{head} node points to \emph{tail}; \emph{head}
(resp. \emph{tail}) is initialized with values $-\infty$ (resp. $+\infty$) that is smaller (resp. greater) than
any other value in the list.

\remove{
In this paper, we consider  the conventional \textsf{set} type exporting
$\lit{insert}$, $\lit{remove}$, and $\lit{contains}$ operations with
standard sequential semantics: $\lit{insert}(v)$
adds $v$ to the set and returns $\true$ if $v$ is not already there,
and returns $\false$ otherwise; $\lit{remove}(v)$ 
drops $v$ from the set and returns $\true$ if $v$ is there,
and returns $\false$ otherwise; and $ \lit{contains}(v)$ returns
$\true$ if and only if $v$ is in the set.
As a baseline for a concurrency analysis of concurrent set implementations, we consider the sequential implementation of the set type (denoted \LL) that uses a \emph{sorted linked list}.  The exact specification and the list-based sequential
implementation of \textsf{set} are presented in Appendix~\ref{app:ll}.
}

\myparagraph{Executions}
An \emph{event} of a process $p_i$ 
is an invocation or response of an operation performed by $p_i$ on a
high-level object (in this paper, a set) implementation, 
or a \emph{primitive} applied by $p_i$ to a base object $b$
along with its response.
A \emph{configuration} specifies the value of each base object and the state of each process.
The \emph{initial configuration} is the configuration in which all 
base objects have their initial values and all processes are in their initial states.
An \emph{execution fragment} is a (finite or infinite) sequence of events.
An \emph{execution} of an implementation $I$ is an execution
fragment where, starting from the initial configuration, each event is
issued according to $I$ and each response of a primitive matches the state of $b$ resulting from all
preceding events.
Let $\alpha|p_i$ denote the subsequence of an execution $\alpha$
restricted to the events of process $p_i$.
Executions $\alpha$ and $\alpha'$ are \emph{equivalent} if for every process
$p_i$, $\alpha|p_i=\alpha'|p_i$.
An operation $\pi$ \emph{precedes} another operation $\pi'$ in an execution
$\alpha$, 
denoted $\pi \rightarrow_{\alpha} \pi'$, 
if the response of $\pi$ occurs before the invocation of $\pi'$ in $\alpha$.
Two operations are \emph{concurrent} if neither precedes
the other. 

An execution is \emph{sequential} if it has no concurrent 
operations. 
An operation is \emph{complete} in $\alpha$ if the invocation event is
followed by a \emph{matching} response; otherwise, it is \emph{incomplete} in $\alpha$.
Execution $\alpha$ is \emph{complete} if every operation is complete in $\alpha$.

\myparagraph{High-level histories and linearizability}
A \emph{high-level history} $\tilde H$ of an execution $\alpha$ is the subsequence of $\alpha$ consisting of all
invocations and responses of 
(high-level) operations.
%
%
A complete high-level history $\tilde H$ is \emph{linearizable} with 
respect to an object type $\tau$ if there exists
a sequential high-level history $S$ equivalent to $\tilde H$ such that
(1) $\rightarrow_{\tilde H}\subseteq \rightarrow_S$ and
(2) \emph{$S$ is consistent with the sequential specification of type $\tau$}.
%
Now a high-level history $\tilde H$ is linearizable if it can be
\emph{completed} (by adding matching responses to a subset of
incomplete operations in $\tilde H$ and removing the rest)
to a linearizable high-level history~\cite{HW90}.
\subsection{Concurrency as admissible schedules of sequential code}
\myparagraph{Schedules}
Informally, a \emph{schedule} 
of a list-based set algorithm specifies the order in which concurrent high-level operations access the list nodes. 
Consider the \emph{sequential}
implementation, \LL, of operations $\lit{insert}$, $\lit{remove}$ and
$\lit{contains}$.
Suppose that we treat this implementation as a \emph{concurrent} one, i.e., simply run it in a concurrent environment, without introducing any synchronization mechanisms, 
and let $\S$ denote the set of the resulting executions, we call them \emph{schedules}. 
Of course, some schedules in $\S$ will not be linearizable.
For example, concurrent inserts operating on the same list nodes may result in ``lost updates'': an inserted element disappears from the list due to a concurrent insert operation.    
But, intuitively, as no synchronization primitives are used, this (incorrect) implementation is as concurrent as it can get.  

We measure the concurrency properties of a linearizable list-based set via its ability to accept all \emph{correct} schedules in $\S$.  
Intuitively, a schedule is correct if it respects the sequential implementation {\LL} \emph{locally}, i.e., no operation in it can distinguish the schedule from a sequential one.   
Furthermore, the schedule must be linearizable, even when we consider its extension in which all update operations are completed and followed with a $\lit{contains}(v)$ for any $v\in\mathbb{Z}$.  
Let us denote this extension of schedule $\sigma$ by $\bar\sigma(v)$. 

\begin{definition}[Correct schedules]
We say that a schedule $\sigma$ of a concurrent list-based set implementation is \emph{locally serializable} 
(with respect to the sequential implementation of list-based set {\LL}) 
if for each of its
operations $\pi$, there exists a sequential schedule $S$ of {\LL}  
such that $\sigma|\pi=S|\pi$. 
We say that a 
schedule is \emph{correct} if 
(1) $\sigma$ is locally serializable (with respect to {\LL}),
(2) for all  $v\in\mathbb{Z}$, $\bar \sigma(v)$ is linearizable (with respect to the \textsf{set} type).
\end{definition}
Note that the last condition is necessary for filtering out schedules with ``lost updates''.
Consider, for example a schedule in which $\lit{insert}(1)$ and $\lit{insert}(2)$ are applied to the initial empty set. 
Imagine that they first both read $\ms{head}$, then both read
$\ms{tail}$, then both perform writes on the $\ms{head.next}$ and complete. 
The resulting schedule is, technically, linearizable and locally serializable but, obviously, not acceptable.  
However, in the schedule, one of the operations, say $\lit{insert}(1)$,
overwrites the effect of the other one.
Thus, if we extend the schedule with a complete execution of
$\lit{contains}(2)$, the only possible response it may give is $\ms{false}$ which obviously does not produce a linearizable high-level history. 

Note also that, as linearizability is a safety property~\cite{Lyn96}, if $\bar \sigma(v)$ is linearizable, $\sigma$ is linearizable too.  
(In the following we omit mentioning \textsf{set} and {\LL} when we talk about local serializability and linearizability.)

\myparagraph{Concurrency-optimality}
A concurrent list-based set generally follows \LL:
every high-level operation, $\lit{insert}$, $\lit{remove}$, or $\lit{contains}$, \emph{reads} the list nodes, one after the other,  until the desired fragment of the list is located. The update operation ($\lit{insert}$ or
$\lit{remove}$) then \emph{writes}, to the $\ms{next}$ field of one of the nodes,
the address of a new node (if it is $\lit{insert}$) or the address
of the node that follows the removed node in the list (if it is
$\lit{remove}$).
Note that the (sequential) write can be implemented using a \textit{CAS} primitive~\cite{harris-set}.

\ignore{
Consider an execution of a concurrent list-based set implementation: this involves reading and writing to the node fields \emph{val} and \emph{next}, as well as reading and modifying \emph{meta-fields} such as \emph{locks} and a boolean that indicates that a node is \emph{marked for deletion}.
%
Consider the subsequence of the execution corresponding to the "sequential" reads, writes (of \emph{val} and \emph{next} fields) and \emph{node creating} events of the sequential implementation \LL.
}

Let $\alpha$ denote an execution of a concurrent implementation of a list-based set.
We define the \emph{schedule $\sigma$ exported by $\alpha$} as the subsequence of $\alpha$ consisting of reads, writes and node creation events (corresponding to the sequential implementation \LL) of operations $\lit{insert}$, $\lit{remove}$ and
$\lit{contains}$ that ``take effect''.
Intuitively, taking effect means that they affect the outcome of some operation. 
The exact way an execution $\alpha$ is mapped to the corresponding schedule $\sigma$ is implementation specific.  
%
\ignore{
Note that taking effect here simply means that, for each operation $\pi$ of the list-based set in $\alpha$ we require $H|\pi=S|\pi$ ($H|\pi$ refers to the restriction of $H$ to events of $\pi$ in $H$ and $S$ is a sequential schedule of \LL).
}

\ignore{
Thus, a schedule is an \emph{equivalence class} of histories that agree on the order of reads, writes, node creation events and high-level operations, but not necessarily on the responses of high-level operations and read events. However, note that in every concurrent implementations designed specifically for the list-based set, every read operation on the base object $b$ returns the value of the latest preceding write (on $b$)~\footnote{This is true of concurrent lists in this paper; however, this claim is not true for list-based set implementations based on generic shared memory abstractions like multi-version transactional memory~\cite{tm-book}}. Thus, for every history, there exists exactly one schedule.
}

\ignore{  
\begin{definition}[Exported schedules]
\label{def:sch}
The schedule exported by an execution $\alpha$ is the
subsequence of $\alpha$ consisting of high-level invocations, high-level
responses, and non-aborted reads and writes on node's \emph{val} and
\emph{next} fields.  
\end{definition}
}
%
%

An implementation $I$ \emph{accepts} a schedule $\sigma$ if there exists an execution of $I$ that exports $\sigma$. 
\begin{definition}[Concurrency-optimality]
An implementation is \emph{concurrency-optimal} if it accepts every correct schedule.   
\end{definition}

\subsection{Concurrency analysis of the Lazy and Harris-Michael Linked Lists}
In this section, we show that even state-of-the-art implementations of the list-based set, namely, the Lazy Linked List
and the Harris-Michael Linked list are suboptimal w.r.t exploiting concurrency. We show that each of these two algorithms rejects some correct schedules of the list-based set.
In order to give an intuition for concurrency analysis, we first describe why the classic \emph{hand-over-hand} locking technique for list-based sets is not concurrency optimal.

\myparagraph{Hand-over-hand locking}
In hand-over-hand locking, each read or write on the list node involves acquiring the \emph{exclusive lock} on the node; the process reads the next field of the node before releasing the lock on the predecessor node in a hand-over-hand manner. 

Consider the following simple example.
Let the list contain values $\{1,2\}$.
For $i=1,2$, let $X_i$ denote the list node used to store $i$. 
Let us now apply a correct schedule in which two operations $\lit{insert}(2)$ and $\lit{insert}(3)$ are applied concurrently: both operations first read $X_1$, then $X_2$ and then $\lit{insert}(2)$ returns $\false$, while $\lit{insert}(3)$ creates a new node and links it to $X_2$.

The schedule is correct, but no execution of the hand-over-hand linked list implementation can export it. 
This is because $\lit{insert}(2)$ must read the value of node $X_2$ prior to releasing the lock on node $X_1$; however $\lit{insert}(3)$ cannot read $X_1$ prior to acquiring the lock on the node. Consequently, there is no execution in which $\lit{insert}(3)$ can read $X_1$ after $\lit{insert}(2)$ reads $X_1$, but before $\lit{insert}(2)$ reads $X_2$.

This example illustrates how concurrent implementations might preclude certain simple classes of interleavings of the steps of the sequential implementation. We now show that the Lazy Linked List and Harris-Michael Linked List are \emph{concurrency sub-optimal}.

\myparagraph{Lazy Linked List}
In this deadlock-free algorithm~\cite{HHL+05}, the list is traversed in the wait-free manner and the locks are taken by update operations only when the desired interval of the list is located.
A $\lit{remove}$ operation first marks a node for \emph{logical} deletion and then \emph{physically} unlinks it from the list.   
To take care of conflicting updates, the locked nodes are \emph{validated}, which involves checking if they are not logically deleted.       
If validation fails, the traversal is repeated. 
The schedule of an execution of this algorithm is naturally derived by considering only the \emph{last} traversal of an operation.

Figure~\ref{fig:ex2} illustrates how the 
post-locking validation strategy employed by the Lazy Linked List makes it concurrency sub-optimal.
As explained in the introduction, the $\lit{insert}$ operation of the Lazy Linked List acquires the lock
on the nodes it writes to, prior to the check of the node's state.
\begin{figure*}[!h]
\scalebox{0.7}{\begin{tikzpicture}
\node (r1) at (1,0) [] {};
\node (r3) at (3,0) [] {};
\node (w1) at (5,0) [] {};

\node (r11) at (2,-1) [] {};
\node (r31) at (7,-1) [] {};

\draw  (1,0)  [black,fill=black, radius=0.1] circle (0.5ex);
\draw  (3,0)  [black,fill=black, radius=0.1] circle (0.5ex);
\draw  (5,0)  [black,fill=black, radius=0.1] circle (0.5ex);

\draw (r1) node [above] {\small {$R(h)$}};
\draw (r3) node [above] {\small {$R(X_1)$}};
\draw (w1) node [above] {\small {$\ms{new}(X_2)$}};

\draw  (2,-1)  [black,fill=black, radius=0.1] circle (0.5ex);
\draw  (7,-1)  [black,fill=black, radius=0.1] circle (0.5ex);
\draw  (9,-1)  [black,fill=black, radius=0.1] circle (0.5ex);

\draw (r11) node [above] {\small {$R(h)$}};
\draw (r31) node [above] {\small {$R(X_1)$}};

\begin{scope}   
\draw [|-,thick] (0,0) node[left] {$\lit{insert}(2)$} to (6,0);
\draw [|-|,thick] (1,-1) node[left] {$\lit{insert}(1)$} to (9,-1) node[right] {$\false$};
\draw [-, dotted] (8,-2) to (8,0.7) node[right] {$E'$};
\draw [-, dotted] (9,-2) to (9,0.7) node[right] {$E$};
\end{scope}
\node[draw,align=left] at (6.5,1.3) {$\lit{insert}(2)$ is incomplete};
\node[draw,align=left] at (12.5,-1.4) {$\lit{insert}(1)$ must acquire\\ the lock on $X_1$ prior to\\ returning $\false$ in $E$};
\node[draw,align=left] at (12.5,0) {$\lit{insert}(2)$ holds the lock\\ on $X_1$ after $E'$};

\end{tikzpicture}}
 \caption{\small{
A schedule rejected by the Lazy Linked List; initial list state is $\{X_1 \}$ that stores value $1$;
$R(X_1)$ refers to reads of both \emph{val} and \emph{next} fields; $\ms{new}(X_2)$ creates a new node storing value $2$}}
\label{fig:ex2}
\end{figure*}
%

One can immediately see that the Lazy Linked List is not concurrency optimal.
%
Indeed, consider the schedule depicted in Figure~\ref{fig:ex2}.
Two operations,  $\lit{insert}(1)$ and $\lit{insert}(2)$ are concurrently applied to the list containing a single node $X_1$ storing value $1$. 
Both operations first read $\lit{h}$, the head of the list, then operation $\lit{insert}(2)$ reads node $X_1$ and creates a new node, $X_2$, storing $2$. 
Immediately after that, operation $\lit{insert}(1)$ reads $X_1$ and returns $\false$. 

The schedule is correct: it is linearizable and locally serializable. 
However, it cannot be accepted by the Lazy Linked List, as $\lit{insert}(1)$ needs a lock on $X_1$ previously acquired by $\lit{insert}(2)$. 
Thus, the implementation is concurrency sub-optimal: an operation may engage in synchronization mechanisms even if it is not going to update the list.    

%
\begin{figure*}[!h]
\scalebox{0.7}{
\begin{tikzpicture}
\node (r1) at (11,0) [] {};
\node (r3) at (15,0) [] {};
\node (w1) at (13,0) [] {};

\node (r11) at (12,1) [] {};
\node (w11) at (14,1) [] {};
\node (r31) at (16,1) [] {};
\node (r41) at (17.5,1) [] {};

\node (r21) at (1,0) [] {};
\node (r23) at (3,0) [] {};
\node (w21) at (5,0) [] {};

\node (r211) at (.5,1) [] {};
\node (r231) at (3,1) [] {};
\node (w211) at (5,1) [] {};

\draw  (11,0)  [black,fill=black, radius=0.1] circle (0.5ex);
\draw  (15,0)  [black,fill=black, radius=0.1] circle (0.5ex);
\draw  (13,0)  [black,fill=black, radius=0.1] circle (0.5ex);

\draw  (12,1)  [black,fill=black, radius=0.1] circle (0.5ex);
\draw  (14,1)  [black,fill=black, radius=0.1] circle (0.5ex);
\draw  (16,1)  [black,fill=black, radius=0.1] circle (0.5ex);
\draw  (17.5,1)  [black,fill=black, radius=0.1] circle (0.5ex);

\draw  (1,0)  [black,fill=black, radius=0.1] circle (0.5ex);
\draw  (3,0)  [black,fill=black, radius=0.1] circle (0.5ex);
\draw  (5,0)  [black,fill=black, radius=0.1] circle (0.5ex);

\draw  (.5,1)  [black,fill=black, radius=0.1] circle (0.5ex);
\draw  (2.5,1)  [black,fill=black, radius=0.1] circle (0.5ex);
\draw  (4.5,1)  [black,fill=black, radius=0.1] circle (0.5ex);

\draw (r1) node [above] {\small {$R(X_2)$}};
\draw (r3) node [above] {\small {$R(X_3)$}};
\draw (w1) node [above] {\small {$W(X_1)$}};

\draw (r11) node [above] {\small {$R(X_2)$}};
\draw (r31) node [above] {\small {$R(X_3)$}};
\draw (w11) node [above] {\small {$W(X_1)$}};
\draw (r41) node [above] {\small {$R(X_4)$}};

\draw (r21) node [above] {\small {$R(h)$}};
\draw (r23) node [above] {\small {$R(X_2)$}};
\draw (w21) node [above] {\small {$W(X_2)$}};

\draw (r211) node [above] {\small {$R(h)$}};
\draw (r231) node [above] {\small {$R(X_2)$}};
\draw (w211) node [above] {\small {$W(h)$}};

\begin{scope}   
\draw [|-|,thick] (0,0) node[left] {$\lit{remove}(2)$} to (6,0) node[right] {$\false$};
\draw [|-|,thick] (0,1) node[left] {$\lit{insert}(1)$} to (6,1) node[right] {$\true$};
\draw [|-|,thick] (10,0) node[left] {$\lit{insert}(3)$} to (16.5,0) node[right] {$\false$};
\draw [|-|,thick] (10,1) node[left] {$\lit{insert}(4)$} to (18,1) node[right] {$\false$};
\end{scope}


\end{tikzpicture}}
 \caption{\small{
A schedule rejected by the Harris-Michael Linked List; the initial state of the list is $\{X_2,X_3,X_4\}$; each $X_i$ 
stores value $i$; note that not all schedules are depicted for succinctness.
}}\label{fig:ex3}%
\end{figure*}
\myparagraph{Harris-Michael Linked List}
Like the Lazy Linked List, the \emph{lock-free} Harris-Michael algorithm (cf. \cite[Chapter 9]{HS12-book})
separates logical deletion of a node from its physical removal (both steps use \emph{CAS} primitives). 
If a CAS associated with logical deletion fails, the operation is restarted. 
Unlike the Lazy Linked List, however, if the physical removal fails (e.g., a concurrent update performed a successful CAS on the preceding node) the operation completes, and unlinking the logically deleted node from the list is then left for a future operation.
Every update operation, as it traverses the list, attempts to physically remove such nodes.
If the attempt fails, the operation is restarted.
The delegation of physical removals to future operations is crucial for lock-freedom: an operation may only be restarted if there is a concurrent operation that took effect, i.e., global progress is made.
But, as we show below, this delegation precludes some legitimate schedules.

Strictly speaking, this algorithm is not locally serializable with respect to the sequential implementation (Algorithm~\ref{alg:lists}).
Indeed, if a $\lit{remove}$ operation completes after logical deletion, we may not be able to map its steps to a write to a $\lit{next}$ field of the preceding node without ``over-writing'' a concurrent update.  
Therefore, for the sake of concurrency analysis, we consider a variant of Algorithm~\ref{alg:lists} in which $\lit{remove}$ operations only remove 
nodes \emph{logically} and physical removals are put to the traversal procedure of future update operations.  
Now to define the schedule incurred by an execution of the algorithm, we consider the read and write steps that are part of the last traversal of an operation, node creation steps by $\lit{insert}$ operations, and successful logical deletions by $\lit{remove}$ operations. 
However, the Harris-Michael Linked List is not concurrency-optimal even with respect to this adjusted sequential specification.
%

Consider the schedule depicted in Figure~\ref{fig:ex3}.
Two operations,  $\lit{insert}(1)$ and 
$\lit{remove}(2)$ are concurrently applied to the list containing three nodes, $X_2$, $X_3$ and $X_4$, storing values $2$, $3$ and $4$, respectively.
Note that operation $\lit{remove}(2)$ 
marks node $X_2$ for deletion
but does not remove it physically by unlinking it from $h$.
(Here we omit steps that are not important for the illustration.)  
Note that so far the schedule is accepted by the Harris-Michael algorithm: an earlier update of $h$ by  operation $\lit{insert}(1)$ causes the corresponding CAS primitive performed on $h$ by $\lit{remove}(2)$ to fail.  

After the operation completes,   we schedule two concurrent operations, $\lit{insert}(4)$ and $\lit{insert}(3)$.
Suppose that the two operations concurrently read $\lit{head}$, $X_1$ and $X_2$. As they both witness $X_2$ to be marked for logical deletion, they both will  try to physically remove it by modifying the $\lit{next}$ field of $X_1$. 
%
We let $\lit{insert}(3)$ to do it first and complete by reading $X_3$ and returning $\false$. 
In the schedule depicted in Figure~\ref{fig:ex3}, $\lit{insert}(4)$ also writes to $X_1$, and then successfully reads $X_3$ and $X_4$, and returns $\false$. 
However, in the execution of the Harris-Michael algorithm, the attempt of $\lit{insert}(4)$ to physically remove $X_2$ will fail, causing it to restart traversing the list from the head. 
Thus, this schedule cannot be accepted.
%

\section{The \name list}
\label{sec:algolist}
In this section, we address the challenges of extracting maximum concurrency from list-based sets and present our \name list.
\subsection{Extracting maximum concurrency from list-based sets}
As we have shown in the previous section, the Lazy Linked List acquires the lock on a node it is about to modify prior to checking the node's state. Thus, it may reject a correct schedule that does not modify the list. 
The schedule rejected by the Harris-Michael Linked List (Figure~\ref{fig:ex3}) is a bit more intricate: it exploits the fact that Harris-Michael List involves \emph{helping} which in turn induces additional synchronization steps leading to rejection of correct schedules.

Deriving a concurrency-optimal list requires introducing pre-locking node validation for the Lazy Linked List along with the combination of the logical deletion from the Harris-Michael Linked List. 
One possible solution for this is to leverage a \emph{versioning} mechanism
(hinted earlier in the TM context~\cite{timebasedtm}) that
allows validating (checking if a node's state has been modified by a concurrent operation) before acquiring the lock. 

However, even seemingly optimal versioning-based techniques for list-based sets fall short of providing concurrency optimality. 
Even though we could present a simple schedule that exhibits sub-optimality of a versioned list, let us consider a somewhat trickier schedule that is not accepted by any known implementation. 
Suppose that the initial state of the list is $\{1,2,3\}$ to which operation $\pi=\lit{remove}(2)$ is applied. %
After $\pi$ reads nodes 
$X_1$ and $X_2$,
a concurrent $\pi'=\lit{remove}(2)$ executes sequentially to completion returning $\true$ following immediately by a sequential execution of $\lit{insert}(2)$ that returns $\true$. Following this, $\pi$ sets $\ms{next}$ pointer of $X_1$ to $X_3$ and returns $\true$ (effectively, removing the just inserted element). This is a correct schedule of the list-based set, however, a versioning-based algorithm that tracks writes to nodes using version numbers would reject this schedule preventing $\pi$ from running to completion and causing it to restart.




This observation inspired our \emph{value-aware try-lock}, used to implement the \name list. This try-lock helps our list to accept schedules similar to the one described above.

\subsection{Value-aware try-lock}
\label{sec:vtrylock}
The class $\lit{Node}$ now contains the fields: (i)~$\lit{val}$ for the value of the node; (ii)~$\lit{next}$ providing a reference to the next node in the list; (iii)~a boolean $\lit{deleted}$ to indicate a node to be marked for deletion and (iv)~a $\lit{lock}$ to indicate a mutex associated with the node.

The value-aware try-lock (Algorithm~\ref{alg:lock})
supports the following operations: 

\begin{enumerate}
    \item[(1)] $\lit{lockNextAt(Node ~node)}$ first acquires the lock on the invoked node, checks if the node is marked for deletion or if the next field does not point to the node passed as an argument, then releases the lock and returns $\false$; otherwise, the operation returns $\true$. 
    
\item[(2)] $\lit{lockNextAtValue(V ~val)}$ acquires the lock on the invoked node, checks if the node is marked for deletion or if the value of the next node is not $\lit{val}$, then releases the lock and returns $\false$; otherwise returns $\true$.
\end{enumerate}

\begin{algorithm*}[ht!]
\caption{Value-aware trylock}
\label{alg:lock}
  \begin{algorithmic}[1]
  	\begin{multicols}{3}
  	{\scriptsize

\Part{Shared variables}{
		\State $\textbf{class Node}$:
  \State ~~~~$\textbf{V}$ val, its value
  \State ~~~~$\textbf{Node}$ next, its reference to the next node
  \State ~~~~bool deleted, a deleted mark
  \State ~~~~$\textbf{Lock}$ lock
}\EndPart

\newpage

\Part{$\textbf{lockNextAt}$(Node node)}{

\State lock.lock()
    \If{ deleted or next $\ne$ node}
      \State lock.unlock()
      \State \textbf{return} false
    \EndIf
    \State \textbf{return} true
}\EndPart

\newpage

\Part{$\textbf{lockNextAtValue}$(V v)}{
\State    lock.lock()
    \If{ deleted or next.v $\ne$ v}
      \State lock.unlock()
      \State \textbf{return} false
      \EndIf
    \State \textbf{return} true
}\EndPart
  
}
	\end{multicols}
  \end{algorithmic}
\end{algorithm*}

\subsection{\name list}
We now describe our \name implementation.
The list is initialized with $2$ nodes: $\ms{head}$ (storing the minimum sentinel value) and $\ms{tail}$
(storing the maximum value),
$\ms{head}.\ms{next}$ stores the pointer to $\ms{tail}$, both $\ms{deleted}$ flags are set to $\ms{false}$.
The pseudo-code is presented in Figure~\ref{alg:concur}.

\myparagraph{Contains} The $\lit{contains}(v)$ algorithm starts from the $\ms{head}$ node and follows $\ms{next}$ pointers until it finds a node with the value that is equal to or bigger than $v$. Then, the algorithm simply compares the value in the found node with $v$.

\myparagraph{Inserting a node} The algorithm of $\lit{insert}(v)$ starts with the traversal (Line~\ref{insert:restart}) to find a pair of nodes $\langle \ms{prev}, \ms{curr} \rangle$ such that $\ms{prev.val}$ is less than $v$ and $\ms{curr.val}$ is equal to or bigger than $v$. The traversal is simple: it starts from $\ms{head}$ and traverses the list in a wait-free manner until it finds the desired nodes. If $\ms{curr.val}$ is equal to $v$ (Line~\ref{insert:equal}) then there is no need to insert. Otherwise, the new node with value $v$ should be between $\ms{prev}$ and $\ms{curr}$. We create a node with value $v$ (Lines~\ref{insert:creation:1}-\ref{insert:creation:2}). Then, the algorithm locks $\ms{prev}$ and checks that it still can insert the node correctly (Line~\ref{insert:check}): $\ms{prev.next}$ still equals to $\ms{curr}$ and $\ms{prev}$ is not marked as deleted. If both of these conditions are satisfied, the new node can be linked. Otherwise, it cannot: the correctness of the algorithm (namely, linearizability) would be violated; so the operation restarts from the traversal (Line~\ref{insert:restart}). Note that to improve the performance, the algorithm starts the traversal not from $\ms{head}$ but from $\ms{prev}$.

\myparagraph{Removing a node}
The algorithm of $\lit{remove}(v)$ follows the lines of $\lit{insert}(v)$: first it finds the desired pair of nodes $\langle prev, curr \rangle$. If $\ms{curr.val}$ is not equal to $v$ then there is nothing to remove (Line~\ref{remove:value}). Otherwise, the algorithm has to remove the node with value $v$. At first, it takes the lock on $\ms{prev}$ and checks two conditions (Line~\ref{remove:check:prev}): $\ms{prev.next.val}$ equals to $v$ and $\ms{prev}$ is not marked as deleted. The first condition 
ensures concurrency-optimality by taking care of the scenario described above: one could have removed and inserted $v$ while the thread was asleep. The second condition is necessary to guarantee correctness, i.e., the node $\ms{next}$ is not linked to \emph{deleted} node, which might result in a ``lost update'' scenario. 
If any of the conditions is violated, the algorithm restarts from Line~\ref{remove:restart}. Then, the algorithm takes the lock on $\ms{curr} = \ms{prev.next}$ and checks a condition $\ms{curr.next}$ equals to $next$ in Line~\ref{remove:check:curr} (note that the second condition is satisfied by the lock on $\ms{prev}$ as $\ms{curr}$ is not marked as deleted). This condition ensures correctness: otherwise, the link $\ms{next}$ to $\ms{prev}$ will be incorrect. If it is not satisfied, the algorithm restarts from Line~\ref{remove:restart}. Afterwards, the algorithm sets $\ms{curr.deleted}$ to $\ms{true}$ (Line~\ref{remove:deleted}) and unlinks $\ms{curr}$ (Line~\ref{remove:unlink}).

\begin{algorithm*}[ht!]
\caption{\name list}
\label{alg:concur}
  \begin{algorithmic}[1]
  	\begin{multicols}{3}
  	{\footnotesize

\Part{Shared variables}{
  \State head.val $\leftarrow -\infty$
  \State tail.val $\leftarrow +\infty$
  \State head.next $\leftarrow$ tail
  \State head.deleted $\leftarrow$ false
  \State tail.deleted $\leftarrow$ false
  \State head.lock $\leftarrow$ new Lock()
  \State tail.lock $\leftarrow$ new Lock()

	}\EndPart
\Statex
\Part{$\lit{contains}$(v)}{
\State curr $\leftarrow$ head
  \While{curr.val < v}
    \State curr $\leftarrow$ curr.next
  \EndWhile
  \State \textbf{return} curr.val = v \label{line:linrl}

}\EndPart

\newpage

\Part{$\lit{waitfreeTraversal}$(v, prev)}{
\If{prev.deleted} \label{line:traversal-begin}
    \State prev $\leftarrow$ head
    \EndIf
  \State curr $\leftarrow$ prev.next
  \While{curr.val < v}
    \State prev $\leftarrow$ curr
    \State curr $\leftarrow$ curr.next
  \EndWhile
  \State \textbf{return} $\langle \text{prev}, \text{curr} \rangle$ \label{line:traversal-end}
  
}\EndPart

\Statex

\Part{$\lit{insert}$(v)}{
\State prev $\leftarrow$ head
  \State $\langle \text{prev, curr} \rangle$ $\leftarrow$ waitfreeTraversal(v, prev)  \label{insert:restart}
  \If{curr.val = v}
    \textbf{return} false
    \EndIf \label{insert:equal}
  \State newNode.val $\leftarrow$ v \label{insert:creation:1}
  \State newNode.next $\leftarrow$ curr \label{insert:creation:2}
  \If{ not prev.lockNextAt(curr)} \label{insert:check}
    \State \textbf{goto} Line~\ref{insert:restart}
  \EndIf
  \State prev.next $\leftarrow$ newNode \label{line:reach}
  \State prev.lock.unlock() \label{line:relinsert}
  \State \textbf{return} true
  
}\EndPart

\newpage

\Part{$\lit{remove}$(v)}{
\State prev $\leftarrow$ head
  \State $\langle \text{prev, curr} \rangle$ $\leftarrow$ waitfreeTraversal(v, prev) \label{remove:restart}
  \If{curr.val $\ne$ v} \label{remove:value}
    \State \textbf{return} false
    \EndIf
  \State next $\leftarrow$ curr.next 
  \If{not prev.lockNextAtValue(v)} \label{remove:check:prev}
    \textbf{goto} Line~\ref{remove:restart}
  \EndIf
  \State curr = prev.next \label{remove:unlink}
  \If{not curr.lockNextAt(next)} \label{remove:check:curr}
    \State prev.unlock()
    \State \textbf{goto} Line~\ref{remove:restart}
  \EndIf
  \State curr.deleted $\leftarrow$ true \label{remove:deleted}
  \State prev.next $\leftarrow$ curr.next \label{remove:unlink}
  \State curr.lock.unlock()
  \State prev.lock.unlock()
  \State \textbf{return} true
}\EndPart

}
	\end{multicols}
  \end{algorithmic}
\end{algorithm*}

\myparagraph{Correctness}
%
We show that the \name list accepts only correct schedules of the list-based set.
In the next section, we show that the \name list accepts \emph{every} correct schedule of the list-based set, thus establishing its concurrency-optimality.
\begin{theorem}
\label{th:lr}
Every schedule of the \name{} list is linearizable with respect to the $\ms{set}$.
\end{theorem}
\begin{proofsketch}
We depict the assignment of the linearization points to show how the partial ordering is constructed that is equivalent to the sequential history of the list-based set; however the full formal proof is delegated 
\iftechrep 
 to the Appendix~\ref{sec:proof}.
\else 
 to the companion technical report~\cite{AGK20}.
\fi For every operation $\pi$, let $\ell_{\pi}$ denote its linearization point in an execution $\alpha$ and $\tilde H$, the corresponding history.

For every $\pi=\lit{insert}(v)$ that returns $\true$ in $\tilde H$, $\ell_{\pi}$ is associated with 
the write event in Line~\ref{line:reach} (rendering the node that
stores $v$ reachable from the \ms{head}); otherwise
$\ell_{\pi}$ is associated with the last $\lit{read}$ of a node's \emph{next} field performed by $\pi$ in $\alpha$. 

For every $\pi=\lit{remove}(v)$ that returns $\true$ in $\tilde H$, $\ell_{\pi}$ is associated with 
the write event in Line~\ref{remove:deleted} (setting the
\ms{deleted} flag of a list's node); otherwise
$\ell_{\pi}$ is associated with the last $\lit{read}$ of a node's \emph{next} field performed by $\pi$ in $\alpha$. 

For $\pi=\lit{contains}(v)$ that returns $\true$, $\ell_{\pi}$
is associated with the last $\lit{read}$ performed by $\pi$.

For $\pi=\lit{contains}(v)$ that returns $\false$ in $H$, $\ell_{\pi}$ is assigned to one of the two events: 
\begin{itemize}
\item
If $\pi$ reads $(X.val\neq v)$ in Line~\ref{line:linrl}, where $X$ is the last node read by $\pi$ in $\alpha$,
$\ell_{\pi}$ is assigned to the read of the \emph{next} field of the node accessed by $\pi$ immediately before $X$.
\item
Alternatively, $\ell_{\pi}$ is defined as follows:
let $\pi_1$ be the $\lit{remove}$ operation that performs the last write
to $X.\ms{deleted}$ (Line~\ref{remove:deleted}) prior to the read of $X$ by $\pi$.
Then, $\ell_{\pi}$ is chosen to be the first event performed by $\pi_1$ immediately after the write to $X.\ms{deleted}$, but prior to
the read of $X$ by $\pi$. Otherwise if no such event of $\pi_1$ exists, then
$\ell_{\pi}$ is the read of $X$ by $\pi$.
\end{itemize}
Given a sequential history ${\tilde S}$ constructed by the above assignment of linearization points, we inductively construct the proof. Let ${\tilde S}^k$ be the prefix of $\tilde S$ consisting of
the first $k$ complete operations. 
We associate each ${\tilde S}^k$ with a set $q^k$ of objects that were
successfully inserted and not subsequently successfully removed in ${\tilde S}^k$.
We show by induction on $k$ (via case by case analysis) that the sequence of state transitions in ${\tilde S}^k$ is consistent with operations' responses in ${\tilde S}^k$ with respect to the \textsf{set} type. 
\end{proofsketch}
\iftechrep
 The Appendix establishes the full proof of the above theorem while Theorem~\ref{th:deadlock} proves that \name list is deadlock-free. Observe that the only nontrivial case to analyse for proving deadlock-freedom is the execution of the update operations. Suppose that an update operation $\pi$ fails to return a matching response after taking infinitely many steps. However, this means that there exists a concurrent $\lit{insert}$ and $\lit{remove}$ that successfully acquires its locks and completes its operation, thus implying progress for at least one correct process.
\else 
 The full proof is deferred to the companion technical report. 
 Observe that the only nontrivial case to analyse for proving deadlock-freedom is the execution of the update operations. Suppose that an update operation $\pi$ fails to return a matching response after taking infinitely many steps. However, this means that there exists a concurrent $\lit{insert}$ or $\lit{remove}$ that successfully acquires its locks and completes its operation, thus implying progress for at least one correct process.
\fi
\begin{theorem}\label{lem:ls}
The \name implementation 
accepts only correct list-based set schedules
locally serializable (wrt {\LL}).
\end{theorem}
\begin{proof}
To show our algorithm is locally serializable, we first remark that every operation
traverses the list starting from the $\ms{head}$ node and reads the
$\ms{next}$ field of a node to locate the subsequent node.  
Before adding a new node to the list (Lines~\ref{insert:creation:1} and \ref{insert:creation:2}),
each $\lit{insert}$ operation initializes the node's $\ms{val}$ and $\ms{next}$ fields, so that at all times the $\ms{next}$ field of a node 
stores a pointer to  an inserted node with a strictly higher  value or
to the $\ms{tail}$ node.
Furthermore, the values stored in the list are integers, every
operation invoked with parameter $v$ eventually locates the node
storing $v$ or a higher value. 
Thus, every sequence of non-aborted 
events (i.e., without prior restart) of every operation $\pi$ is finite.
Hence, there exists a sequence of insert
operations $S_0$, such that $S_0\cdot\sigma|\pi$ is a sequential
schedule of {\LL}.         
\end{proof}
%
%

\myparagraph{Concurrency-optimality}
We prove that the \name accepts every correct interleaving of the sequential code.
The goal is to show that any finite schedule rejected by our algorithm  is not correct. 
Recall that a correct schedule $\sigma$ is locally serializable and, when extended with all its update operations completed and $\lit{contains}(v)$, for    any $v\in \mathbb{Z}$, 
we obtain a linearizable schedule.

Note that given a correct schedule, we can define the  
contents of the list from the order of the schedule's write operations. For each
node that has ever been created in this schedule, we derive the resulting state of its
\textit{next} field from the last write in the schedule. 
Since in a correct schedule each new node is first created and then
linked to the list, we can reconstruct the \emph{state of the list} by iteratively traversing it, starting from the $\ms{head}$.    

%
\begin{theorem}[Optimality]
\label{th:lrelaxed}
The \name implementation accepts all correct schedules. 
\end{theorem}
\begin{proof}
Let $\tilde \sigma$ be a schedule rejected by {\name{}}. Let $\sigma$ be the longest prefix of $\tilde \sigma$ that was accepted by {\name{}}.
For every update operation $\pi$ in $\sigma$, $\sigma|\pi$ is defined to be the subsequence
$\sigma|\pi$ consisting of the reads, writes and node creation events from the last
invocation of the function $\lit{waitfree-traversal}$ by $\pi$ extended by the execution in which $\pi=\lit{insert}$ (and resp. $\pi=\lit{remove}$) executes contiguously without restarting in Line~\ref{insert:check} (and resp. Line~\ref{remove:check:prev} and \ref{remove:check:curr}).
For every $\pi=\lit{contains}$ in $\sigma$, $\sigma|\pi$ is the subsequence
$\sigma|\pi$ consisting of the reads and writes on a node's \emph{val} and \emph{next} fields.

We argue that any schedule \emph{rejected} by our algorithm is  
not observably correct.
More precisely, an operation restarts a fragment
of it execution (in Lines~\ref{insert:check}, \ref{remove:check:curr}, \ref{remove:check:prev}) only if extending it with a read or a write
on $\ms{next}$ or $\ms{val}$ field would result in a schedule that is
not observably correct.

We consider two cases:
\begin{itemize}
    \item 
We first observe that if a node is logically deleted
(Line~\ref{remove:deleted}), then its next write renders the node
unreachable from the \ms{head} node. 
Thus, an update operation $\pi$ partially restarts because of reading a logically
deleted node (in function $\lit{lockNextAt}$) only if it is concurrent with a
\lit{remove} operation which, when completed would physically remove
the node addressed by $\pi$ at the end of its traversal. 
It is easy to see that regardless of what this operation $\pi$ is
($\lit{insert}(v)$ or $\lit{remove}(v)$), if we complete it in turn
and then extend the resulting schedule with $\lit{contains}(v)$, the
effect of $\pi$ will not be seen and the schedule will not be
linearizable.

\item Similarly, let $\sigma$ be the schedule up to the prefix where an update operation $\pi$ partially restarts in Lines~\ref{remove:check:curr}. The update operation restarts because it fails in grabbing a lock on one of the nodes it is about to modify, i.e.,
$\pi$ is concurrent with another update operating on the same node.  
By completing both $\pi$ and the concurrent update, we obtain an extension of the 
schedule in which one of the updates is ``lost'', so that its
extension with some $\lit{contains}(v)$ will not be linearizable.                 
\end{itemize}
\end{proof}

\section{Experimental evaluation}
\label{sec:eval}
\myparagraph{Experimental setup}
In this section, we compare the performance of our solution to two state-of-the-art 
list-based set algorithms written in different languages (Java and C++) and on two multicore machines from different manufacturers:
\begin{itemize}
    \item A 4-socket Intel Xeon Gold 6150 2.7 GHz server (Intel) with 18 cores per socket (yielding 72 cores in total), 512 Gb of RAM, running Debian 9.9. This machine has OpenJDK 11.0.3. 
    \item A 4-socket AMD Opteron 6276 2.3 GHz server (AMD) with 16 cores per socket (yielding 64 cores in total), running Ubuntu 14.04. This machine has OpenJDK 1.8.0\_222.
\end{itemize}

\myparagraph{Concurrent list implementations}
We compared our \name algorithm (VBL) to the lock-based Lazy Linked List (Lazy)~\cite{HHL+05} and Harris-Michael's non-blocking 
list (Harris-Michael)~\cite{harris-set,michael-set} with its wait-free and RTTI optimization suggested by Heller et al.~\cite{HHL+05} 
using the Synchrobench benchmark suite~\cite{Gra15}. 
To compare these algorithms on the same ground we primarily used Java as it is agnostic of the underlying set up.
The evaluation of the C++ implementations of these algorithms is 
\iftechrep 
 presented in Appendix~\ref{sec:eval:cpp}.
\else 
 deferred to the companion technical report~\cite{AGK20}.
\fi
The code of the implementations 
is part of Synchrobench at \url{https://github.com/gramoli/synchrobench}.

\myparagraph{Experimental methodology}
To evaluate the performance we considered the following parameters:
\begin{itemize}
    \item \textbf{Workloads.} Each workload distribution is characterized by the percent $x\%$ of update operations. This means that the list will be requested to make $(100 - x)\%$ of $\lit{contains}$ calls, $x/2\%$ of $\lit{insert}$ calls and $x/2\%$ of $\lit{remove}$ calls. We considered three different workload distribution: $0\%$, $20\%$, and $100\%$. Percentages $0\%$ and $100\%$ were chosen as the extreme workloads, while $20\%$ update ratio corresponds to the standard load on databases. Each operation $\lit{contains}$, $\lit{insert}$, and $\lit{remove}$ chooses its argument uniformly at random from the fixed key range.
    \item \textbf{List size.} On the workloads described above, the size of the list depends on the range from which the operations take the arguments. Under the described workload the size of the list is approximately equal to the half of the key range. We consider four different key ranges: $50$, $200$, $2 \cdot 10^3$, and $2 \cdot 10^4$. To ensure consistent results we pre-populated the list: each element is present with probability $\nicefrac{1}{2}$.
    \item \textbf{Degree of contention.} This depends on the number of cores in a machine. We take enough points to reason about the behavior of the curves.
\end{itemize}
\begin{figure*}[!h]
\scalebox{.8}{
\includegraphics[width=1\linewidth]{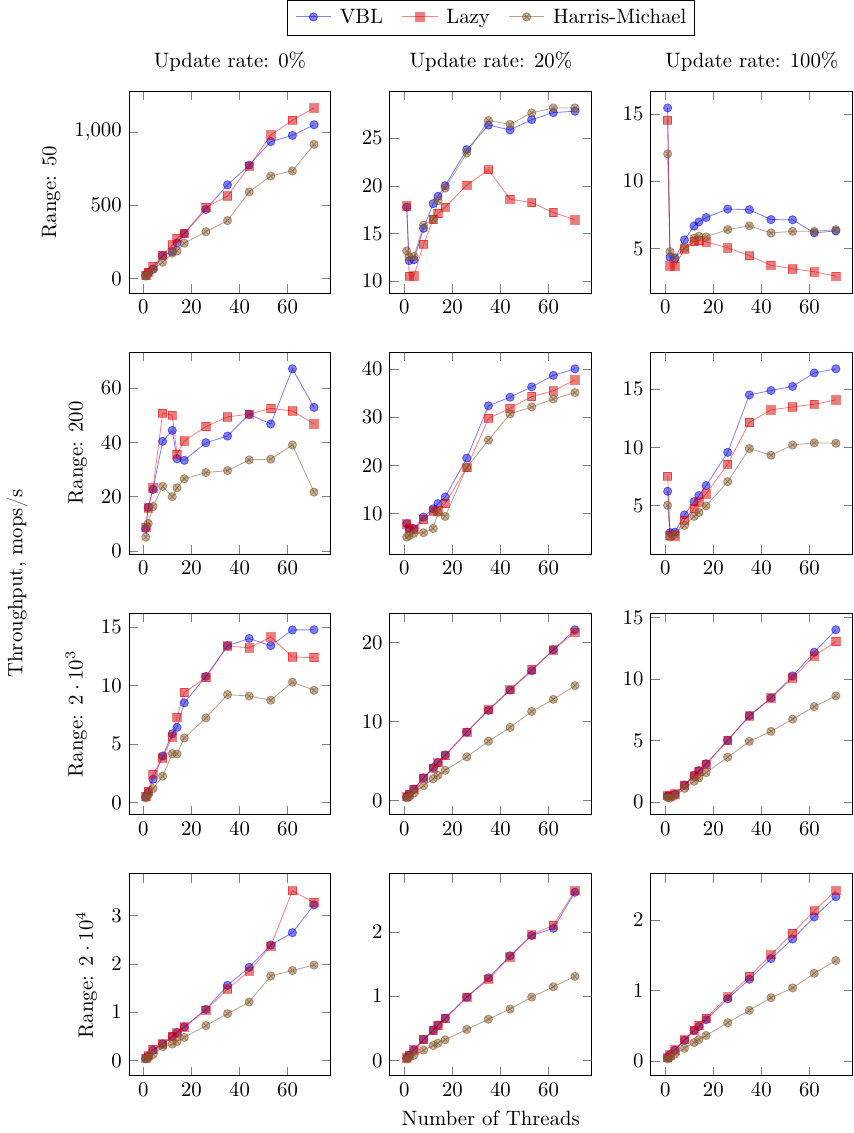}
}
 \caption{\small{Evaluation on Intel
}}\label{sfig:results:intel}%
\end{figure*}

\myparagraph{Results} We run experiments for each workload 5 times for 5 seconds with a warm-up of 5 seconds. 
Figure~\ref{sfig:results:intel} (and resp. Figure~\ref{sfig:results:amd}) contains the results of executions on Intel (and resp. AMD) machine.
Our new list algorithm outperforms both Harris-Michael's and the Lazy Linked List algorithms, and 
remains scalable except for the situation with very high contention, i.e., high update ratio with small range. We find this behavior normal at least in our case, since the processes contend to get the cache-lines in exclusive mode and this traffic becomes the dominant factor of performance in the execution.

\myparagraph{Comparison against Harris-Michael} 
Harris-Michael's algorithm in general scales well and performs 
well under high contention. 
Even though the three algorithms 
feature the wait-free $\lit{contains}$, 
our original implementation of the Harris-Michael's $\lit{contains}$ was slower than the other two.
The reason is the extra indirection needed when reading the $\ms{next}$ pointer in the 
combined \emph{pointer-plus-boolean} structure.
To avoid reading an extra field when fetching the Java $\lit{AtomicMarkableReference}$
we implemented the run-time type identification (RTTI) variant with two subclasses that inherit from a 
parent node class and that represent the marked and unmarked 
states of the node as previously suggested~\cite{HHL+05}. 
This optimization requires, on the one hand, that a $\lit{remove}$ casts the subclass instance to the parent class to create a corresponding node in the marked state. 
It allows, on the other hand, the traversal to simply check the mark of each node by simply 
invoking $\lit{instanceof}$ on it to check the subclass the node instantiates. 
As we see, Harris-Michael's algorithm has very efficient updates because it only 
uses \emph{CAS}, however it spends much longer on list traversals. 

\myparagraph{Comparison against the Lazy Linked List}  
The Lazy Linked List has almost the same performance as our algorithm under low contention 
because both share the same wait-free list traversal with 
zero overhead (as the sequential code does) and for the updates, 
when there is no interference from concurrent operations, 
the difference between our pre-locking-validation and Heller et al.'s post-locking-validation becomes negligible. The difference raises however as the contention appears.
The Lazy Linked List performance drops significantly due to its intense lock competition 
(as briefly explained in Section~\ref{sec:intro}). 
By contrast, there are several features in our implementation that reduce significantly the amount of contention on the locks. 
We observed a tremendous increase in execution time for the Lazy Linked List because of the 
contention on locks. 
\begin{figure*}[!h]
\scalebox{.8}{
\includegraphics[width=1\linewidth]{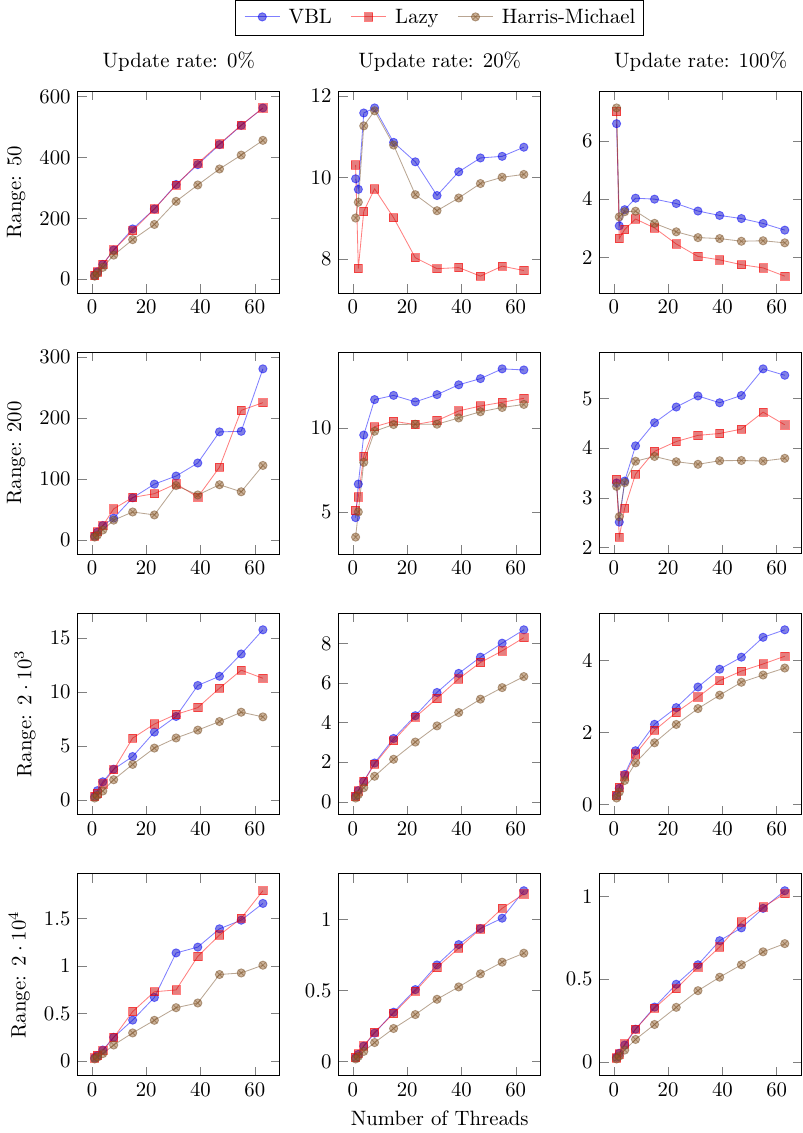}
}
 \caption{\small{Evaluation on AMD
}}\label{sfig:results:amd}%
\end{figure*}

\section{Discussion}\label{sec:conclusion}
\myparagraph{List-based sets}
Heller \emph{et al.}~\cite{HHL+05} proposed the Lazy Linked List
and mentioned the option of validating prior to locking, and using a single lock within an $\lit{insert}$.
One of the reasons why our implementation is faster than the Lazy Linked List is 
the use of a novel value-aware try-lock mechanism
that allows validating before acquiring the lock.

Harris~\cite{harris-set} proposed a non-blocking linked list algorithm
that 
splits the removal of a node into two atomic steps: a logical deletion
that marks the node and a physical removal 
that unlinks the node from the list.
Michael~\cite{michael-set} proposed advanced memory reclamation algorithms for the algorithm of Harris.
In our implementation, we 
rely on Java's garbage collector for memory reclamation~\cite{hotspot}. 
We believe that our implementation could outperform Michael's variant for the same reason it 
outperforms Harris' one because it does not 
combine the logical deletion mark with the next pointer of a node but separates metadata 
(logical deletion and versions) from the structural data (check~\cite{HS12-book} for variants of these list-based sets).

Fomitchev and Ruppert~\cite{FR04} proposed a lock-free linked list where nodes have a backlink field that allows to backtrack 
in the list in case a conflict is detected instead of restarting 
from the beginning of the list. Its \textit{contains} operation also helps remove marked nodes from the list.  
Gibson and Gramoli~\cite{GG15} proposed the \emph{selfish linked list}, as a more efficient variant  of this approach with the same amortized complexity, relying on wait-free \textit{contains} operations. 
These algorithms are, however, not concurrency-optimal: schedule constructions similar to those outlined for the Harris-Michael and Lazy linked lists apply here.

\myparagraph{Concurrency metrics}
%
Sets of accepted schedules are commonly used as a
metric of concurrency provided by a shared-memory
implementation.
For static database transactions, 
Kung and Papadimitriou~\cite{KP79} use the metric to 
capture the parallelism of a locking scheme.
While acknowledging that the metric is theoretical, they 
insist that it may
have ``practical significance as
well, if the schedulers in question have relatively small
scheduling times as compared with waiting and execution
times.'' 
Herlihy~\cite{Her90} employed the metric from \cite{KP79} to compare various
optimistic and pessimistic synchronization techniques using
commutativity~\cite{Wei88} 
of operations constituting high-level transactions.   
A synchronization technique is implicitly considered in~\cite{Her90} as highly
concurrent, namely ``optimal'',
if no other technique accepts more schedules. 
In contrast to~\cite{KP79,Her90}, we focus here on a \emph{dynamic} model where the scheduler cannot use the prior knowledge of all the shared addresses to be accessed.

Optimal concurrency can also be seen as a variant of metrics like \emph{permissiveness}~\cite{GHS08-permissiveness} and \emph{input acceptance}~\cite{GHF10} defined for transactional memory. 
The concurrency framework considered in this paper though is independent of the synchronization technique and, thus, more general.

Concurrent interleavings of \emph{sequential} code has been used as a base-line for evaluating performance of search data structures~\cite{DGT15}.
Defining \emph{optimal concurrency} as the ability of accepting \emph{all} correct interleavings has been originally proposed and used to compare concurrency properties of optimistic 
and pessimistic techniques
in~\cite{GKR16}. 

\myparagraph{The case for concurrency-optimal data structures}
Intuitively, the ability of an implementation to successfully process
interleaving steps of concurrent threads is an appealing property that
should be met by performance gains.

In this paper, we support this intuition by presenting a
concurrency-optimal list-based set that outperforms (less concurrent) state-of-the-art algorithms.        
Does the claim also hold for other data structures? 
%
We believe that generalizations of linked lists, such as skip-lists or tree-based dictionaries, may allow for optimizations similar to the ones
proposed in this paper. 
The recently proposed concurrency-optimal tree-based dictionary~\cite{bst-co} justifies this belief.  
This work presents the opportunity to construct a rigorous methodology for deriving concurrency-optimal data structures that also  perform well.

Also, there is an interesting intermingling  between progress conditions, concurrency properties, and performance. For example, the Harris-Michael algorithm is superior with respect to both the Lazy Linked List and \name{ }in terms of progress (lock-freedom is a strictly stronger progress condition than deadlock-freedom). However, as we observe, this superiority does not necessarily imply better performance. 
Improving concurrency seems to provide more performance benefits than boosting liveness.
Relating concurrency and progress in concurrent data structures remains an interesting research direction. 



\bibliography{references}

\iftechrep
 \appendix
 \subsection{Proof of linearizability of \name list}
\label{sec:proof}
\begin{theorem}
\label{th:lr}
Every schedule of \name list is linearizable with respect to the $\ms{set}$.
\end{theorem}
\begin{proof}
Let $\alpha$ be a finite execution of \name implementation and $<_\alpha$
denote the total-order on events in $\alpha$.
For the sake of the proof, we assume that $\alpha$ starts with an artificial
sequential execution of an insert operation $\pi_0$ that inserts $\ms{tail}$ and sets $\ms{head}.\ms{next}=\ms{tail}$. 
Let $H$ be the high-level history exported by $\alpha$.
Since \emph{linearizability} is a \emph{safety}
property~\cite{Lyn96}, 
it is sufficient for us to prove that every \emph{finite} high-level history $H$ is linearizable.

\vspace{0.8em}\noindent\textit{Completions.}
We obtain a completion $\tilde H$ of $H$ as follows.
The invocation of an incomplete $\lit{contains}$ operation is discarded.
The invocation of an incomplete $\pi=\lit{remove}$
operation that has not performed the write in Line~\ref{remove:deleted} 
is discarded; otherwise, it is completed with response $\true$.
The invocation of an incomplete $\pi=\lit{insert}$
operation that has not performed the write in Line~\ref{insert:creation:1} 
is discarded; otherwise, it is completed with response $\true$.

\vspace{0.8em}\noindent\textit{Linearization points.}
We obtain a sequential high-level history $\tilde S$ equivalent to $\tilde H$ by associating a linearization point $\ell_{\pi}$ 
with each operation $\pi$ as follows.
%

For every $\pi=\lit{insert}(v)$ that returns $\true$ in $\tilde H$, $\ell_{\pi}$ is associated with 
the write event in Line~\ref{line:reach} (rendering the node that
stores $v$ reachable from the \ms{head}); otherwise
$\ell_{\pi}$ is associated with the last $\lit{read}$ of a node's \emph{next} field performed by $\pi$ in $\alpha$. 

For every $\pi=\lit{remove}(v)$ that returns $\true$ in $\tilde H$, $\ell_{\pi}$ is associated with 
the write event in Line~\ref{remove:deleted} (setting the
\ms{deleted} flag of a list's node); otherwise
$\ell_{\pi}$ is associated with the last $\lit{read}$ of a node's \emph{next} field performed by $\pi$ in $\alpha$. 

For $\pi=\lit{contains}(v)$ that returns $\true$, $\ell_{\pi}$
is associated with the last $\lit{read}$ performed by $\pi$.

For $\pi=\lit{contains}(v)$ that returns $\false$ in $H$, $\ell_{\pi}$ is assigned to one of the two events: 
\begin{itemize}
\item
if $\pi$ reads $(X.val\neq v)$ in Line~\ref{line:linrl}, where $X$ is the last node read by $\pi$ in $\alpha$,
$\ell_{\pi}$ is assigned to the read of the \emph{next} field of the node accessed by $\pi$ immediately before $X$
\item
Alternatively, $\ell_{\pi}$ is defined as follows:
let $\pi_1$ be the $\lit{remove}$ operation that performs the last write
to $X.\ms{deleted}$ (Line~\ref{remove:deleted}) prior to the read of $X$ by $\pi$.
Then, $\ell_{\pi}$ is chosen to be the first event performed by $\pi_1$ immediately after the write to $X.\ms{deleted}$, but prior
the read of $X$ by $\pi$. Otherwise if no such event of $\pi_1$ exists, then
$\ell_{\pi}$ is the read of $X$ by $\pi$.
\end{itemize}

Since linearization points are chosen within the intervals of 
operations performed in $\alpha$, for any two operations
$\pi_i$ and $\pi_j$ in ${\tilde H}$, if $\pi_i \rightarrow_{\tilde H}
\pi_j$, then $\pi_i \rightarrow_{\tilde S} \pi_j$.

Let ${\tilde S}^k$ be the prefix of $\tilde S$ consisting of
the first $k$ complete operations. 
We associate each ${\tilde S}^k$ with a set $q^k$ of objects that were
successfully inserted and not subsequently successfully removed in ${\tilde S}^k$.
We show by induction on $k$ (via case by case analysis) that the sequence of state transitions in
${\tilde S}^k$ is consistent with operations' responses in ${\tilde S}^k$ with respect to the \textit{set} type. 
The base case $k=1$ is trivial: the \textit{tail} node containing
$+\infty$ is successfully inserted.
Suppose that ${\tilde S}^k$ is consistent with the \emph{set} type and
let $\pi_1$ with argument $v\in \mathbb{Z}$ and response $r_{\pi_{1}}$
be the last operation of ${\tilde S}^{k+1}$.  
We want to show that $(q^k,\pi_1,q^{k+1},r_{\pi_{1}})$ is consistent with the \textit{set} type. 


\vspace{2mm}\noindent (1)
Let $\pi_1=\lit{insert}(v)$ return $\true$ in ${\tilde S}^{k+1}$. 
We show below that each preceding $\pi_2=\lit{insert}(v)$ 
returning $\true$ is followed by $\lit{remove}(v)$ 
returning $\true$, such that
$\pi_2 \rightarrow_{{\tilde S}^{k+1}} \lit{remove}(v) \rightarrow_{{\tilde S}^{k+1}} \pi_1$.
Suppose the opposite. 
Observe that $\pi_1$ performs its penultimate $\lit{read}$ on a node $X$ 
that stores a value $v'<v$ and the last read is performed on a node that stores a value $v''>v$. 
By construction of $\tilde S$, $\pi_1$ is linearized at the write on node $X$ in Line~\ref{line:reach}.
Observe that $\pi_2$ must also perform a $\lit{write}$ to the node $X$ 
(otherwise it is easy to see that one of $\pi_1$ or $\pi_2$ would return $\false$).
By assumption, the write to $X$ in shared-memory by $\pi_2$ in Line~\ref{line:reach} 
precedes the corresponding write to $X$ in
shared-memory by $\pi_1$. 
But $\pi_1$ can return $\true$ from the \emph{lock} performed in Line~\ref{insert:check} only after 
$\pi_2$ releases the lock on $X_1$ by performing the event in Line~\ref{line:relinsert}.
Thus, $\pi_1$ could not have returned $\true$---a contradiction.

Let $\pi_1=\lit{insert}(v)$ return $\false$ in ${\tilde S}^{k+1}$.
We show that there exists a preceding $\pi_2=\lit{insert}(v)$ returning 
$\true$ that is not followed by  $\pi_3=\lit{remove}(v)$ returning
$\true$, such that
$\pi_2 \rightarrow_{{\tilde S}^{k+1}} \pi_3 \rightarrow_{{\tilde S}^{k+1}} \pi_1$. 
Suppose that such a $\pi_2$ does not exist. Thus, $\pi_1$ must perform
its last $\lit{read}$ on a node $X$ that stores value $v''>v$, acquire the lock on $X$ (Line~\ref{{insert:check}})
and return $\true$---a contradiction to the assumption that $\pi_1$ returned $\false$.

It is easy to verify that the conjunction of the above two claims proves that 
$\forall q\in Q$; $\forall v\in \mathbb{Z}$, ${\tilde S}^{k+1}$ satisfies $(q,\lit{insert}(v),q \cup \{v\},(v \not\in q))$.

\vspace{2mm}\noindent (2)
If $\pi_1=\lit{remove}(v)$, similar arguments as applied to $\lit{insert}(v)$ prove that 
$\forall q\in Q$; $\forall v\in \mathbb{Z}$, ${\tilde S}^{k+1}$ satisfies $(q,\lit{remove}(v),q \setminus \{v\},(v \in q))$.

\vspace{2mm}\noindent (3)
We now consider the case of $\lit{contains}(v)$.
Let $\pi_1=\lit{contains}(v)$ return $\true$ in ${\tilde S}^{k+1}$. 
We show that there exists $\pi_2=\lit{insert}(v)$ returning \emph{true} 
that is not followed by any $\lit{remove}(v)$ returning \emph{true}, such that
$\pi_2 \rightarrow_{{\tilde S}^{k+1}} \lit{remove}(v) \rightarrow_{{\tilde S}^{k+1}} \pi_1$.

Recall that $\pi_1$ is linearized at the last $\lit{read}$ of a node, say $X$, performed by $\pi$ when 
$\pi$ reads the $\ms{deleted}$ field
of $X$ to be $\false$ (Line~\ref{line:linrl}). 
By the \name algorithm, 
there exists
$\pi_2=\lit{insert}(v)$ such that $\pi_2 \rightarrow_{{\tilde S}^{k+1}} \pi_1$ 
(let $\pi_2$ be the latest such operation).
Suppose that there exists a $\lit{remove}(v)$ that returns
\emph{true}, such that
$\pi_2 \rightarrow_{{\tilde S}^{k+1}} \lit{remove}(v) \rightarrow_{{\tilde S}^{k+1}} \pi_1$.
Thus, $\lit{remove}(v)$ performs the write event in Line~\ref{remove:deleted} prior to the 
read of $X.\ms{deleted}$ by $\pi_1$. But then $\pi_1$ must read $X.\ms{deleted}$ to be $\true$
and return $\false$---a contradiction.

Now, let $\pi_1=\lit{contains}(v)$ return $\false$ in ${\tilde S}^{k+1}$. 
Thus, (1) there exists a $\pi_2=\lit{remove}(v)$ returning
\emph{true} that is not  
followed by any $\lit{insert}(v)$ returning \emph{true},
such that 
$\pi_2 \rightarrow_{{\tilde S}^{k+1}} \lit{insert}(v) \rightarrow_{{\tilde S}^{k+1}} \pi_1$, or
(2) there does not exist any $\lit{insert}(v)$ returning $\true$ such that
$\lit{insert}(v) \rightarrow_{{\tilde S}^{k+1}} \pi_1$.

We consider two cases:
\begin{itemize}
\item 
Suppose that $\pi_1$ reads $(X.value \neq v)$, where $X$ is the last node read by $\pi_1$ in $\alpha$.
Thus, $\ell_{\pi_1}$ is assigned to the read of the \emph{next} field of the node, say $X'$ accessed by $\pi_1$ 
immediately before $X$.
Assume by contradiction that there exists
$\pi_2=\lit{insert}(v)$ that returns \emph{true} such that 
there does not exist any $\lit{remove}(v)$ that returns \emph{true}; 
$\pi_2 \rightarrow_{{\tilde S}^{k+1}} \lit{remove}(v) \rightarrow_{{\tilde S}^{k+1}} \lit{contains}(v)$.
But then $\pi_1$ must read $(X.value = v)$ and return $\true$---contradiction.
\item
Suppose that $\pi_1$ reads $(X.value = v)$ and $X.\ms{deleted}$ is $\true$ in Line~\ref{line:linrl}.
Clearly, there exists a $\pi_2=\lit{remove}(v)$ that is concurrent to $\pi_1$ and returns $\true$ in ${\tilde H}$.
By the assignment of linearization points, $\ell_{\pi_1}$ is assigned to the first event performed
by $\pi_2$ immediately after the write to $X.\ms{deleted}$, but prior to the read of $X.\ms{deleted}$ by $\pi_1$, where $X$
is the last node read by $\pi_1$. 

We consider two cases: \\
(1) Suppose that some such event of $\pi_2$ exists.
We claim that there does not exist any $\pi_3=\lit{insert}(v)$ that returns $\true$ such that
$\pi_2 \rightarrow_{{\tilde S}^{k+1}} \pi_3 \rightarrow_{{\tilde S}^{k+1}} \pi_1$.
Any such $\pi_2$ must acquire the \name lock on $X'$, 
the node read by $\pi_1$ immediately prior to $X$. 
Since $\pi_1$ reads $(X.value = v)$ and $X.\ms{deleted}$ to be $\true$, $\pi_2$ must also acquire the lock
on $X'$. By our assumption, $\ell_{\pi_2} \rightarrow_{{\tilde S}^{k+1}} \ell_{\pi_3}$.
Thus, $\pi_3$ acquires the \name lock on $X'$ only after $\pi_2$ releases it.
But we linearize $\pi_1$ prior to $\ell_{\pi_3}$---a contradiction to our assumption that $\ell_{\pi_3} <_{\alpha} \ell_{\pi_1}$.\\
(2)
Otherwise, if no such event of $\pi_2$ exists, $\ell_{\pi_1}$ is chosen as
the read of $X.\ms{deleted}$ by $\pi_1$. Since $\pi_2$ does not release the value-aware lock on $X'$
prior to the read of $X.\ms{deleted}$ by $\pi_1$, there does not exist any $\lit{insert}(v)$ that returns $\true$ such that 
$ \lit{insert}(v) \rightarrow_{{\tilde S}^{k+1}} \pi_1$. Now, by the assignment of linearization points,
$\pi_2 \rightarrow_{{\tilde S}^{k+1}} \pi_1$.
\end{itemize}
Thus, inductively, the sequence of state transitions in ${\tilde S}$
satisfies the sequential specification of the \textit{set} type. 
\end{proof}
\begin{theorem}
\label{th:deadlock}
The \name list provides deadlock-freedom.
\end{theorem}
\begin{proof}
Suppose that an update operation $\pi$ fails to return a matching response after taking infinitely many steps.
Intuitively, $\pi$ fails to do so because (1)
it fails to acquire the lock on the $\ms{prev}$ node in Line~\ref{insert:check} for $\lit{insert}$, or (2) it is returned $\false$ from the call to 
$\lit{lockNextAtValue}$ for $\lit{remove}$, or
or (3) it fails to acquire the lock on the $\ms{curr}$ node during Line~\ref{remove:check:curr} by $\lit{remove}$.

In Case (1), there exists a concurrent $\lit{remove}$ operation that is executing the code
in its critical section and must eventually return the response $\true$.
In Case (2), if $\pi$ returns $\false$ from the call to $\lit{lockNextAtValue}$, then some concurrent update operation has indeed succeeded to acquire the lock and enters its critical section if its an $\lit{insert}$. Otherwise if this concurrent operation is a $\lit{remove}$, we are in Case (3), i.e., $\lit{remove}$ does not return from the call to  $\lit{curr.lockNextAt}$ in Line~\ref{remove:check:curr}.
Thus, some other update operation has acquired the lock on the $\ms{curr}$ node of this $\lit{remove}$. Note that we can
iteratively extend this execution, but the last such correct process performing an update operation must eventually enter the critical
section and return a matching response.

The schedule of $\alpha$ is defined as the sub-sequence of its events consisting of, for every operation, read operations of the last traversal (lines~\ref{line:traversal-begin}-\ref{line:traversal-end}), node creation events (line~\ref{insert:creation:1}), and writes to the $\lit{next}$ fields (lines~\ref{line:reach} and~\ref{remove:unlink}).   
One can easily see that the sequence of these events follows, for each operation in $\alpha$, the sequential specification of the linked list.
Thus, the schedule is locally serializable. 

Therefore, all schedules accepted by our algorithm are correct.
\end{proof}

 \subsection{Evaluation in C++}
\label{sec:eval:cpp}
In this section, we present the evaluation of the performance of our algorithm on Intel machine (see Section~\ref{sec:eval}) with the code written in C++. We compiled the code with MinGW 6.3.0, \texttt{-O2} flag, and linked it with \texttt{TCmalloc} allocator.

We compare our VBL algorithm with Lazy Linked List and original Harris-Michael algorithm with the only change: we do not support memory reclamation in any algorithm, i.e., the allocated nodes are never reclaimed. The code is available by the following link \url{https://cutt.ly/icdcs2021_677}.

We use the same workloads as in the Section~\ref{sec:eval}. The results are shown at Figure~\ref{fig:results:intel:cpp}.

\begin{figure*}[!h]
\scalebox{.9}{
\includegraphics[width=1\linewidth]{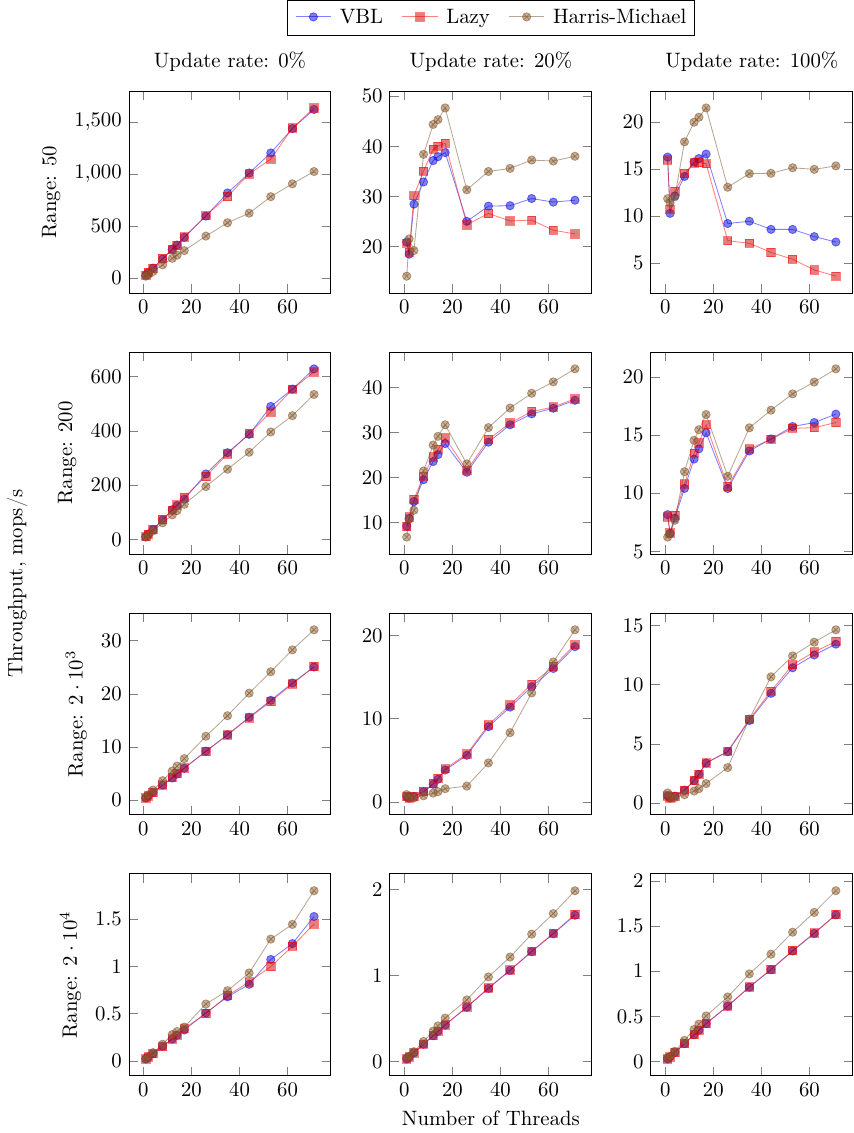}
}
\caption{\small{Evaluation on Intel in C++}}
\label{fig:results:intel:cpp}
\end{figure*}

\myparagraph{Comparison against the Lazy Linked List}
On workloads with high-contention, i.e., short list (range $50$ and range $200$), high update rate ($20\%$ and $100\%$), and high number of processes ($19-72$, out of one socket), VBL outperforms Lazy. On all other workloads VBL performs similarly to Lazy. Thus, we can state that the application of the notion of concurrency-optimality improves the performance of lock-based algorithms.

\myparagraph{Comparison against Harris-Michael.} It can be seen that on workloads with high-contention Harris-Michael significantly outperforms VBL and Lazy. This situation is the opposite to what happens with the algorithms written in Java. It happens because in C++ we can store the next pointer and the deleted flag inside one field and there is no necessity to follow two references in order to get the next pointer. On all other workloads Harris-Michael outperforms VBL and Lazy a little bit or even performs worse.

The only thing that we cannot unequivocally explain is why Harris-Michael unambiguously outperforms VBL and Lazy on contains-only workloads with high range when the \texttt{contains} operation in VBL and Lazy is implemented as a simple while loop. We relate this behaviour to the fact that the compiler could have done his job better on Harris-Michael and the function \texttt{contains} is aligned in the address-space more friendly to the operating system.
\fi 

\end{document}